# Efficiency Limit of Transition Metal Dichalcogenide Solar Cells

Koosha Nassiri Nazif,[1†] Frederick U. Nitta,[1,2†] Alwin Daus,[1,3] Krishna C. Saraswat,[1,2] and Eric Pop[1,2*]

[1]Dept. of Electrical Engineering, Stanford University, Stanford, CA 94305, USA

[2]Dept. of Materials Science and Engineering, Stanford University, Stanford, CA 94305, USA

[3]RWTH Aachen University, Aachen, 52074, Germany

[†]These authors contributed equally.

*corresponding author email: epop@stanford.edu

*Abstract* — **Transition metal dichalcogenides (TMDs) show great promise as absorber materials in high-specific-power (i.e. high-power-per-weight) solar cells, due to their high optical absorption, desirable band gaps, and self-passivated surfaces. However, the ultimate performance limits of TMD solar cells remain unknown today. Here, we establish the efficiency limits of multilayer $MoS_2$, $MoSe_2$, $WS_2$, and $WSe_2$ solar cells under AM 1.5 G illumination as a function of TMD film thickness and material quality. We use an extended version of the detailed balance method which includes Auger and defect-assisted Shockley-Reed-Hall recombination mechanisms in addition to radiative losses, calculated from measured optical absorption spectra. We demonstrate that single-junction solar cells with TMD films as thin as 50 nm could in practice achieve up to 25% power conversion efficiency with the currently available material quality, making them an excellent choice for high-specific-power photovoltaics.**

## I. INTRODUCTION

Transition metal dichalcogenides (TMDs) have recently received growing interest in high-specific-power (i.e. high-power-per-weight) photovoltaics where light weight and high power conversion efficiency are strongly desired[1–4]. TMD materials such as $MoS_2$ and $WSe_2$ have high optical absorption coefficients, desirable band gaps for use in single-junction and tandem solar cells (~1.0 to 2.5 eV), and self-passivated surfaces free of dangling bonds, enabling high performance even for ultrathin absorber layers on the order of 100 nm[2,4–6]. Recently, ultrathin TMD solar cells reached high specific power on par with established thin-film solar technologies cadmium telluride, copper indium gallium selenide, amorphous silicon and III-Vs, with the potential to achieve unprecedented power per weight, 10 times higher than commercialized solar cell technologies[4].

Moreover, adopting ultrathin TMD absorber layers minimizes material utilization, therefore helping with sustainable material use and cost reduction. In addition, the chemical and mechanical stability of TMDs[7] promises reliable and long-lasting performance similar to silicon solar panels, while their biocompatibility[8] allows usage in wearable and implantable electronics in contact with the human body. At the same time, rapid developments in the nanoelectronics industry related to TMD growth and device fabrication[9–12] pave the way for low-cost mass production of TMD solar cells, similar to how silicon solar cells benefited in their early days from the developments made in the microelectronics industry. It is therefore timely to determine the ultimate performance limits of TMD solar cells, illustrating their potential for next-generation solar cell technology which could be realized after sufficient optimization.

In this work, we establish the fundamental performance limits of single-junction solar cells made of multilayer (bulk) $MoS_2$, $MoSe_2$, $WS_2$, and $WSe_2$ absorber films with a realistic analysis based on the Tiedje-Yablonovitch model originally developed for silicon solar cells[13]. This detailed balance model uses material-specific optical absorption data and includes radiative and Auger recombination as well as free carrier absorption, providing material-specific, thickness-dependent performance limits, as opposed to Shockley-Queisser models[2,14], which assume that absorptance steps from zero to unity at the band gap energy. We improve our predictions beyond the Tiedje-Yablonovitch model by incorporating defect-assisted Shockley-Read-Hall (SRH) recombination, finding thickness-dependent efficiency limits for various material quality levels. As a consequence, we find that up to 25% power conversion efficiency is achievable in ultrathin (~50 nm) single-junction TMD solar cells even with existing material quality, corresponding to ~10 times higher power per weight than commercialized solar cell technologies[4]. This already renders TMD photovoltaics an excellent choice for high-specific-power applications such as autonomous drones, electric vehicles, Internet-of-Things devices, and wearable electronics, which are rapidly growing and soon becoming an integral part of our daily life experience.

## II. METHODS

The extended detailed balance method developed by Tiedje *et al.* (known as the Tiedje-Yablonovitch model) is used as the basis for this study[13]. The model was originally developed for silicon solar cells to provide an accurate estimate of their efficiency limits by incorporating the optical absorption characteristics of silicon, radiative and Auger recombination, and free carrier absorption. In this study, we go beyond the Tiedje-Yablonovitch model and investigate the effect of material quality on the solar cell performance by including defect-assisted SRH recombination, as detailed in **Supplementary Note 1**. This comprehensive model, for the first time, provides efficiency limits of single-junction, multilayer TMD solar cells as a function of TMD film thickness and material quality.

To mimic optimal light trapping, a rectangular slab of semiconducting TMD ($MoS_2$, $MoSe_2$, $WS_2$, and $WSe_2$) with perfect anti-reflection (zero reflection) at the front surface and perfect (unity) reflection at the back surface is considered (**Fig. 1**). The illumination (AM 1.5 G spectrum with one-sun intensity) includes both direct and diffuse sunlight over a full $2\pi$-steradian acceptance angle, appropriate for a non-tracking flat solar panel. The surfaces are assumed to be textured non-specular (Lambertian), e.g. created by etching, leading to randomized light and angle-independent absorption. This gives a mean path length of $4n^2L$ for light rays in the semiconductor[13,15], where $L$ is the TMD film thickness (here, from 5 nm to 1 μm) and $n$ is the semiconductor refractive index. Because $n$ is relatively constant across all wavelengths of interest[16], we use the $n$ value at the band gap energy. The operating temperature is assumed to be 300 K.

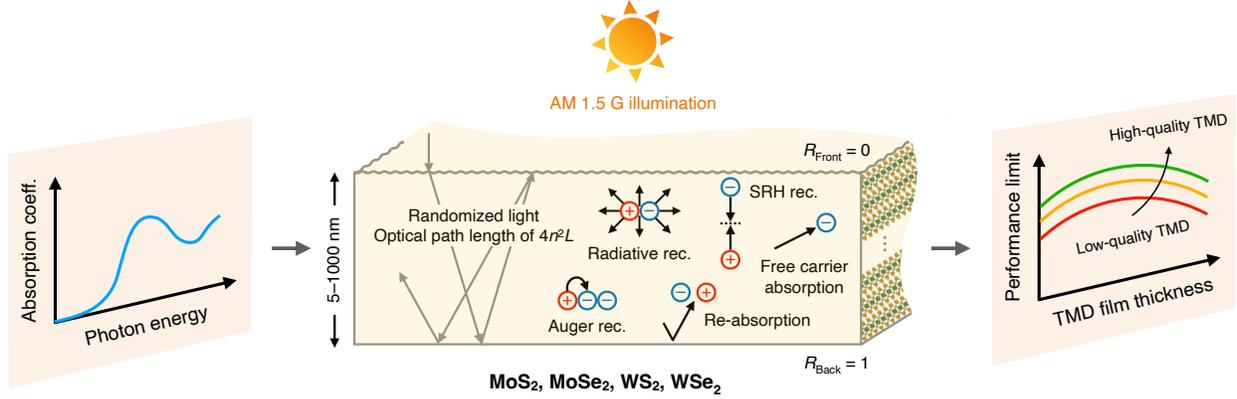

**Figure 1. Modeling setup** showing solar cell geometry, multilayer TMDs modeled, incident sunlight, absorption assumptions, recombination mechanisms, input optical absorption coefficient spectrum, and output thickness- and material quality-dependent performance limit. *R*, reflection; *L*, TMD film thickness; *n*, refractive index.

**Table I | Modeling parameters for bulk $MoS_2$, $MoSe_2$, $WS_2$, and $WSe_2$**[16–18]. **Effective masses and densities of states are appropriately averaged over the in-plane and cross-plane TMD components.**

| Material | $MoS_2$ | $MoSe_2$ | $WS_2$ | $WSe_2$ |
| --- | --- | --- | --- | --- |
| Band gap, $E_G$ (eV) | 1.27 | 1.16 | 1.36 | 1.29 |
| Refractive index at $E_G$, $n$ | 4.48 | 3.67 | 4.68 | 4.63 |
| Effective electron mass, $m_e^*$ | $0.71 m_e$ | $0.64 m_e$ | $0.63 m_e$ | $m_e$ |
| Effective hole mass, $m_h^*$ | $0.84 m_e$ | $0.97 m_e$ | $0.84 m_e$ | $0.59 m_e$ |
| Effective conduction band density of states, $N_C$ (cm$^{-3}$) | $1.50 \times 10^{19}$ | $1.29 \times 10^{19}$ | $1.26 \times 10^{19}$ | $2.51 \times 10^{19}$ |
| Effective valence band density of states, $N_V$ (cm$^{-3}$) | $1.93 \times 10^{19}$ | $2.40 \times 10^{19}$ | $1.93 \times 10^{19}$ | $1.14 \times 10^{19}$ |
| Intrinsic carrier concentration, $n_i$ (cm$^{-3}$) | $3.70 \times 10^{8}$ | $3.20 \times 10^{9}$ | $5.93 \times 10^{7}$ | $2.49 \times 10^{8}$ |
| Auger recombination coefficient (cm$^6$ s$^{-1}$) | $10^{-29.7}$ | $10^{-29.3}$ | $10^{-30.0}$ | $10^{-29.7}$ |

Radiative, Auger, and SRH recombination mechanisms are all considered, as described in **Supplementary Note 1**. Measured optical absorption coefficient spectra of bulk TMDs[16] (**Supplementary Fig. 1**) are used to accurately calculate both absorptance and the radiative losses, and to extract the optical band gap of TMD films using the Tauc method[19] (**Supplementary Fig. 2**). SRH lifetime ($\tau_{SRH}$) is varied from 1 ns to infinity (the case in Tiedje-Yablonovitch model) to determine efficiency limits at various material quality levels. Auger coefficients are extrapolated from Auger coefficient–band gap charts in the literature[18]. A summary of modeling parameters is listed in **Table I**. Intrinsic or lightly-doped TMDs are considered such that hole and electron densities are equal under illumination. At low doping densities and small absorber thicknesses, free carrier absorption is negligible in ultrathin absorber layers[20]. We therefore exclude free carrier absorption from our analysis. Given the low exciton binding energy in bulk TMDs[21,22], it is assumed that all excitons are dissociated into electrons and holes (e.g., by the electric field present across the TMD film). The model outputs the performance limits of the solar cell, particularly power conversion efficiency limit, as a function of TMD film thickness and material quality.

## III. RESULTS AND DISCUSSION

To highlight the unusually high light absorption in thin TMD films, we calculate the spectral absorptance of multilayer $MoS_2$, $MoSe_2$, $WS_2$, and $WSe_2$ at 5, 10, 20, 50, 100 and 1000 nm film thickness (**Fig. 2**) using their measured optical absorption coefficient spectra[16] (**Supplementary Fig. 1**). Due to their large absorption coefficients and refractive indices, all these TMDs exhibit significant light absorptance even in ultrathin films of 5 nm thickness (**Fig. 2**), four orders of magnitude thinner than conventional silicon solar cell absorber layers (~200 μm). As the thickness approaches 1000 nm, the absorptance approaches the simplified step-function assumption in the Shockley-Queisser model[14], with exponential Urbach tails[23] arising from exciton-phonon and exciton-defect interactions in TMDs[24]. The absorptance peaks are mainly attributed to the A and B excitons in these materials[25] (**Supplementary Fig. 1**).

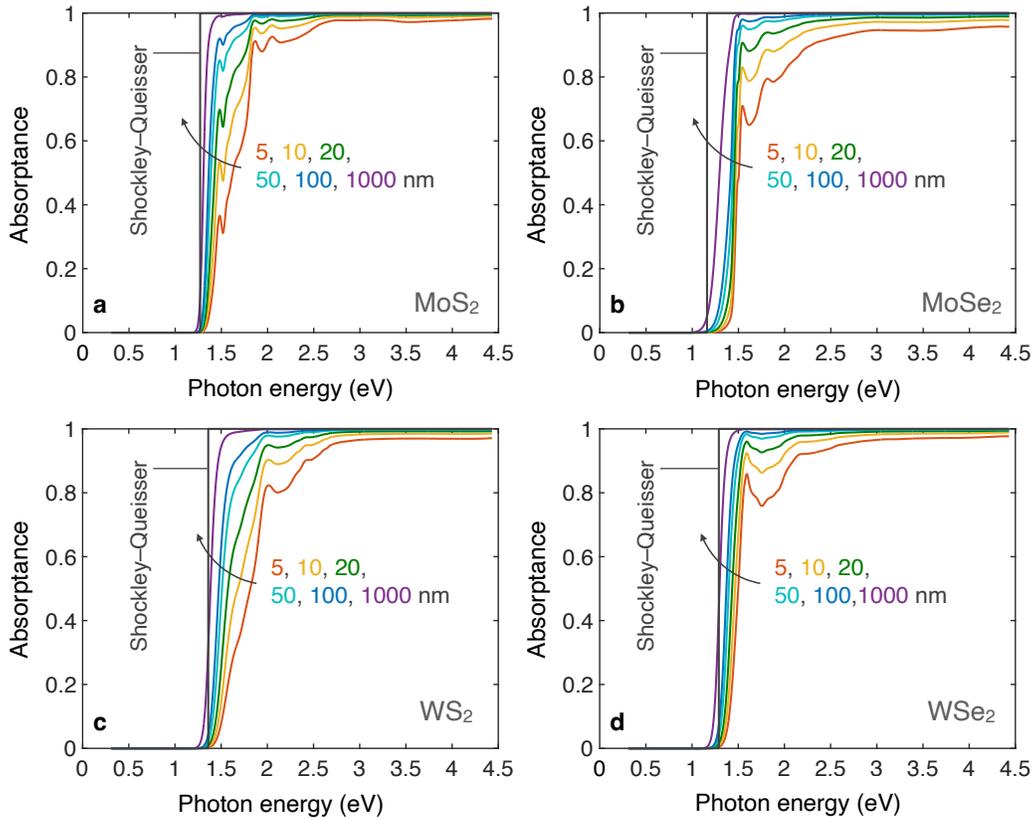

**Figure 2. Spectral absorptance** of **a)** $MoS_2$, **b)** $MoSe_2$, **c)** $WS_2$, and **d)** $WSe_2$ at various thicknesses between 5 and 1000 nm, along with the step-function Shockley-Queisser model. The Tauc band gap[19] of the materials is used for the Shockley-Queisser model (see Supplementary Fig. 2).

The relative absorptance of the four TMDs can be explained by their respective absorption coefficient spectra (**Supplementary Fig. 1**), particularly in the 1–2.5 eV range, and by their refractive indices (**Table I**). Selenides ($MoSe_2$ and $WSe_2$) have larger absorption coefficients than the sulfides ($MoS_2$ and $WS_2$), leading to steeper and higher absorptance in the 1–2.5 eV regime, beyond which near-unity absorptance is reached in all four TMDs, even in ultrathin films of only 5 nm thickness. $WSe_2$ has a larger refractive index, and thus longer optical path length ($4n^2L$) compared to $MoSe_2$, leading to the highest absorptance among the four TMDs. On the other hand, $WS_2$ has the smallest absorption coefficient in the 1–2.5 eV range, with a refractive index comparable to $MoS_2$, therefore making it the least light-absorptive of the four TMDs.

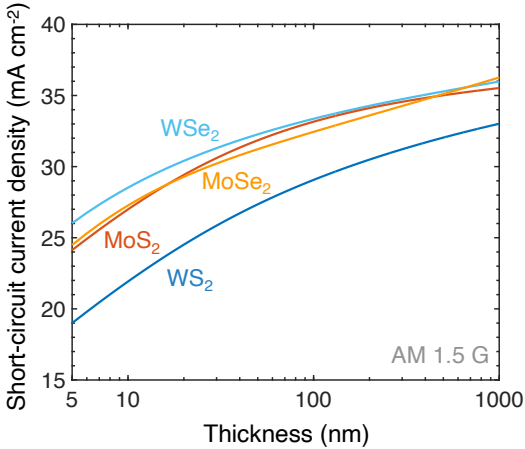

**Figure 3. Short-circuit current densities** of $MoS_2$, $MoSe_2$, $WS_2$, and $WSe_2$ solar cells as a function of the TMD (absorber) film thickness, at 300 K and AM 1.5 G illumination.

**Figure 3** shows the calculated short-circuit current density ($J_{SC}$) of TMD solar cells as a function of the TMD (absorber) film thickness. As expected from their exceptional light absorption characteristics, all TMDs can achieve high $J_{SC}$ even at small thicknesses. Absorptance and therefore $J_{SC}$ increase with increasing TMD film thickness. Radiative, Auger and SRH recombination do not affect the $J_{SC}$ limits within the thickness and SRH lifetime ranges modeled here, particularly due to the low carrier density at zero bias in the intrinsic or lightly-doped TMDs assumed (see **Supplementary Note 1** for more details). In the simple detailed balance Shockley-Queisser model, semiconductors with smaller band gap exhibit higher $J_{SC}$, because they absorb a larger portion of the AM 1.5 G spectrum, with photon energies above their band gap. However, as evident in **Fig. 3**, this is not necessarily true with the extended Tiedje-Yablonovitch method, where absorptance is determined by optical absorption coefficient and refractive index. We observe that $J_{SC}$ follows the same trend as absorptance, with $WSe_2$ and $WS_2$ showing the highest and lowest $J_{SC}$, respectively, and $MoSe_2$ and $MoS_2$ in between.

We also note a change in slope of the $J_{SC}$ trends in **Fig. 3**, where $J_{SC}$ increases more strongly at smaller thicknesses, but then rises at a lower rate in thicker films. The initial steeper $J_{SC}$ increase with thickness can be explained by the noticeable absorptance enhancement (**Fig. 2**) in the ~1.5–2.5 eV regime as the TMD film thickness approaches ~20 nm (MoSe$_2$ and WSe$_2$) to ~50 nm (MoS$_2$ and WS$_2$). Beyond these thicknesses, the absorptance improvement in the ~1.5–2.5 eV region is less prominent. Note that the absorption threshold shifts by approximately 0.2 eV to lower energies as the film thickness increases from 5 nm to 1000 nm (see **Fig. 2**). This shift is the main driver for the continued, yet gentler $J_{SC}$ increase beyond ~20 nm (~50 nm) in MoSe$_2$ and WSe$_2$ (MoS$_2$ and WS$_2$). The absorption threshold shift is more pronounced in MoSe$_2$ (**Fig. 2**), enabling it to achieve larger $J_{SC}$ than MoS$_2$ and WSe$_2$ at large thicknesses beyond ~600 nm.

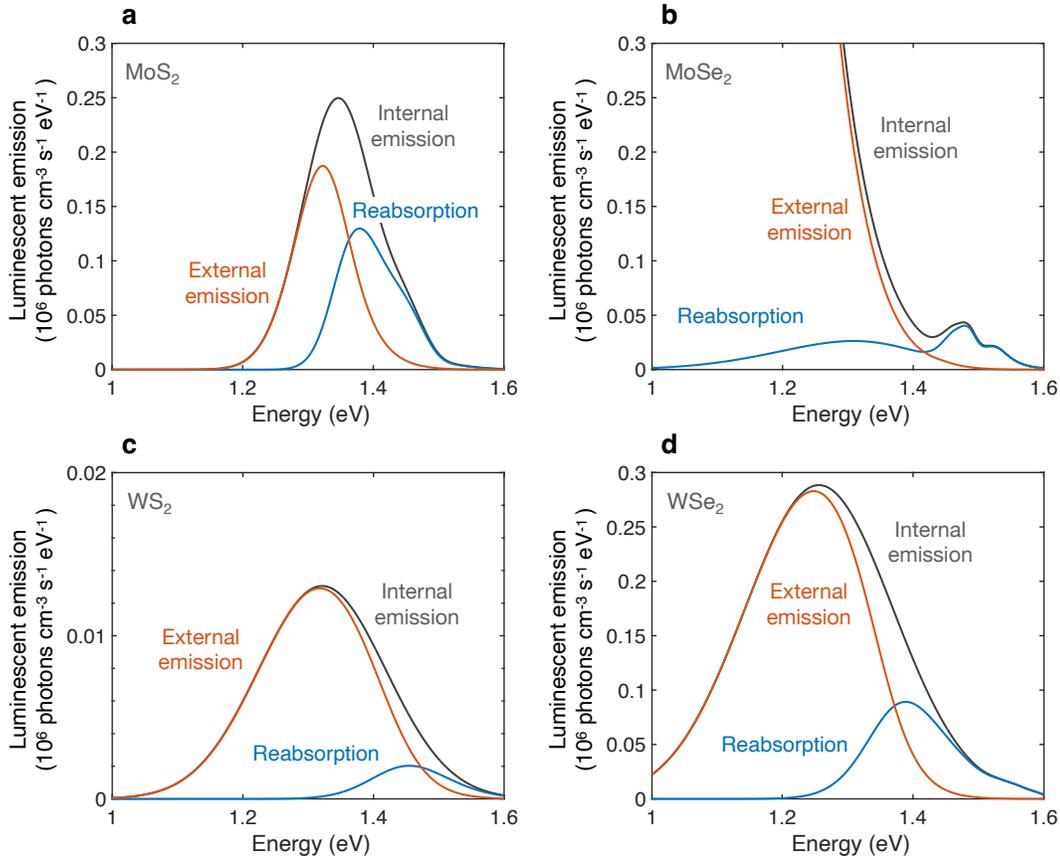

**Figure 4. Spectral dependence of the luminescent emission rates** for a 100 nm-thick film of **a)** MoS$_2$, **b)** MoSe$_2$, **c)** WS$_2$, and **d)** WSe$_2$ in thermal equilibrium at 300 K. Note the vertical axis for WS$_2$ (panel **c**) is smaller than the vertical axes of the other three panels.

Examining the radiative losses, **Fig. 4** shows the spectral dependence of the luminescent emission rates for 100 nm-thick TMD films in thermal equilibrium at 300 K, as described in **Supplementary Note 2**. One can observe that reabsorption is almost equally probable as external emission in MoS$_2$. Moreover, the radiative loss is primarily from the low-energy photons, which have higher absorption depth and therefore lower probability of being reabsorbed into the TMD film. A similar behavior is observed in MoSe$_2$, WS$_2$, and WSe$_2$, with external emission occurring at longer wavelengths (higher energy photons) and reabsorption taking place at shorter wavelengths (lower energy photons). The magnitude of emission rates varies among the four TMDs due to the difference in their absorption coefficients and refractive indices. As detailed in **Supplementary Note 2**, at equilibrium, the internal emission rate is proportional to the absorption coefficient and the square of refractive index. For example, MoSe$_2$ has noticeably higher absorption coefficient than other TMDs in the 1–1.3 eV range due to of its smaller band gap (1.16 eV), leading to the highest internal emission rates. The opposite holds true for WS$_2$, which has the largest band gap (1.36 eV) and smallest absorption coefficient in the entire 1–1.6 eV range (**Supplementary Fig. 2**). The reabsorption rate is equal to the product of the internal emission rate and the absorptance (**Fig. 2**). Finally, the external emission rate, in the absence of free carrier absorption, is the difference between the rates of internal emission and reabsorption.

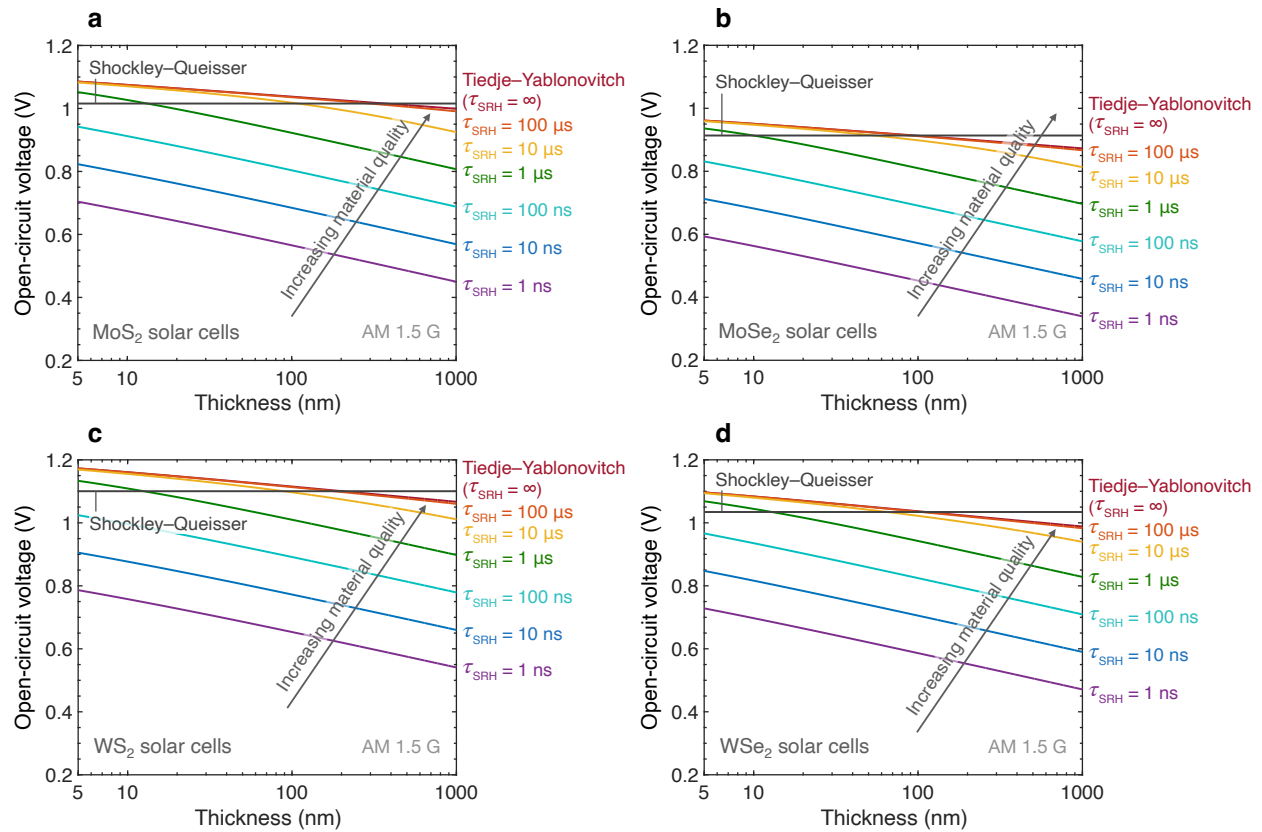

**Figure 5. Open-circuit voltage** of **a)** MoS$_2$, **b)** MoSe$_2$, **c)** WS$_2$, and **d)** WSe$_2$ solar cells as a function of TMD film thickness and material quality (SRH lifetime, $\tau_{SRH}$), at 300 K and AM 1.5 G solar illumination.

The calculated open-circuit voltage ($V_{OC}$) of TMD solar cells as a function of TMD film thickness and material quality (SRH lifetime, $\tau_{SRH}$) is shown in **Fig. 5**, along with the estimate from the Shockley-Queisser model. Infinite SRH lifetime corresponds to the Tiedje-Yablonovitch model where defect-assisted SRH recombination is excluded. In **Fig. 5**, we demonstrate the noticeable effect of material quality on $V_{OC}$ for $\tau_{SRH}$ smaller than 100 μs where SRH recombination starts to dominate the $V_{OC}$ loss. Among these four TMDs, we note that WS$_2$ has the largest $V_{OC}$ for any given $\tau_{SRH}$, due to its largest band gap among the TMDs investigated here. Incidentally, an SRH lifetime up to ~611 ns has also been reported[3] for multilayer $WS_2$, although lifetimes for this and other TMDs are all expected to increase as the material quality improves. Applied to the materials studied here, such an SRH lifetime would lead to a $V_{OC}$ limit between 0.8 to 1 V in 100 nm-thick TMD solar cells.

We note that for thin TMD films, our model estimates a larger $V_{OC}$ limit than the simpler Shockley-Queisser model for $\tau_{SRH}$ larger than 1 μs. This is due to our inclusion of measured optical absorption spectra. As can be seen in **Fig. 2**, the absorption threshold depends on thickness and occurs at higher photon energies in thinner TMD films, yielding an effectively larger band gap than the simple Shockley-Queisser model, a discrepancy which becomes greater for thinner films. This highlights the inadequacy of the step-function absorption assumption in the Shockley-Queisser model, where only one threshold (band gap) energy is used for all film thicknesses.

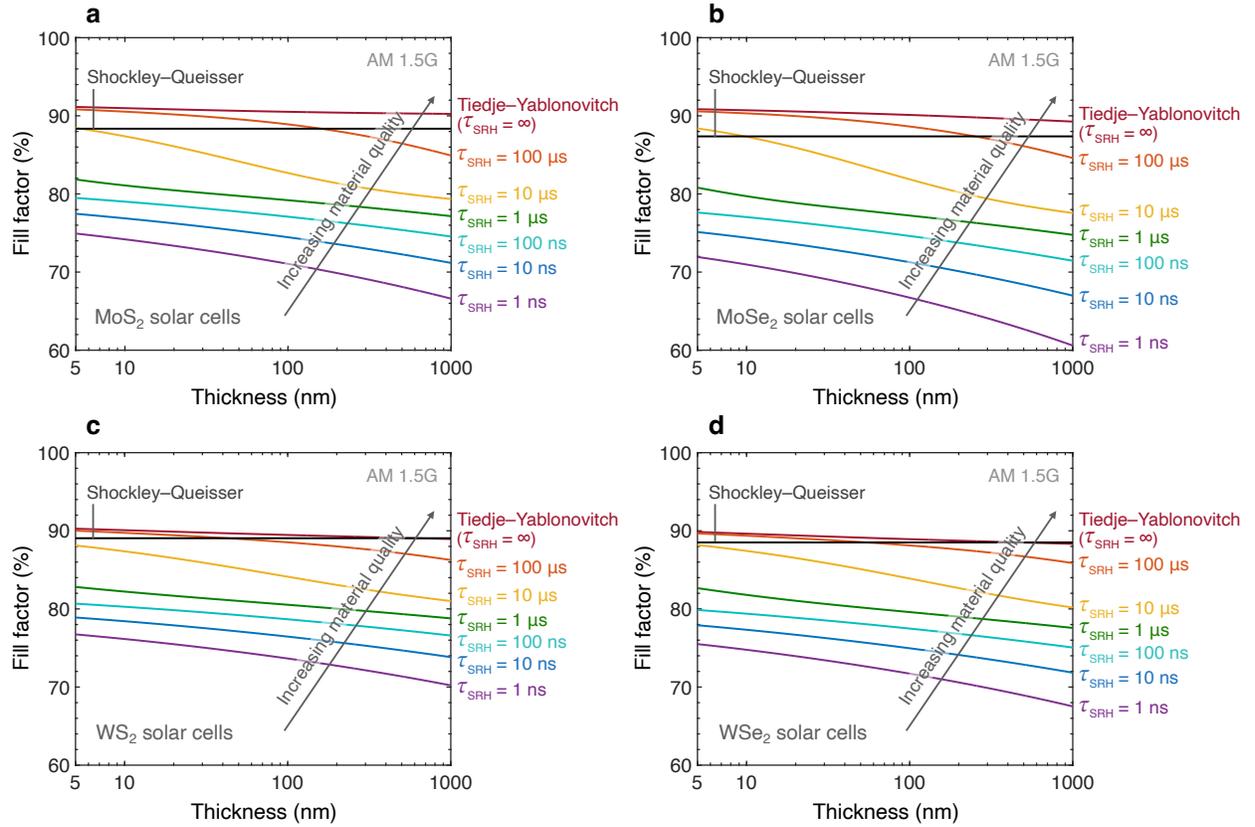

**Figure 6. Fill factor** of **a)** $MoS_2$, **b)** $MoSe_2$, **c)** $WS_2$, and **d)** $WSe_2$ solar cells as a function of TMD film thickness and material quality (SRH lifetime, $\tau_{SRH}$), at 300 K and AM 1.5 G solar illumination.

We also investigate the effect of TMD film thickness and material quality on the fill factor of the four types of TMD solar cells in **Fig. 6**. It is well-known that the larger the $V_{OC}$, the higher the fill factor of the solar cell [26]. Therefore, $WS_2$, having the largest band gap and $V_{OC}$, shows the highest fill factor, and $MoSe_2$, which has the smallest band gap and $V_{OC}$, exhibits the lowest fill factor among the four TMDs. The fill factor dependence on $V_{OC}$ also explains why the fill factor decreases with increasing thickness and decreasing material quality – following the same trend as $V_{OC}$ (**Fig. 5**). Studies also show that the closer the solar cell (diode) ideality factor to unity, the higher the fill factor[26], which explains the higher fill factor in the absence of SRH recombination ($\tau_{SRH} \rightarrow \infty$) compared to the case where $\tau_{SRH} = 100$ μs even though the two have essentially the same $V_{OC}$. Dominant SRH recombination (i.e. $\tau_{SRH} < 100$ μs) leads to an ideality factor of 2 at high-level injection[27,28], which is the case here since the semiconductor is assumed intrinsic or lightly-doped, whereas dominant Auger recombination gives an ideality factor of 2/3[27,28], leading to higher fill factor.

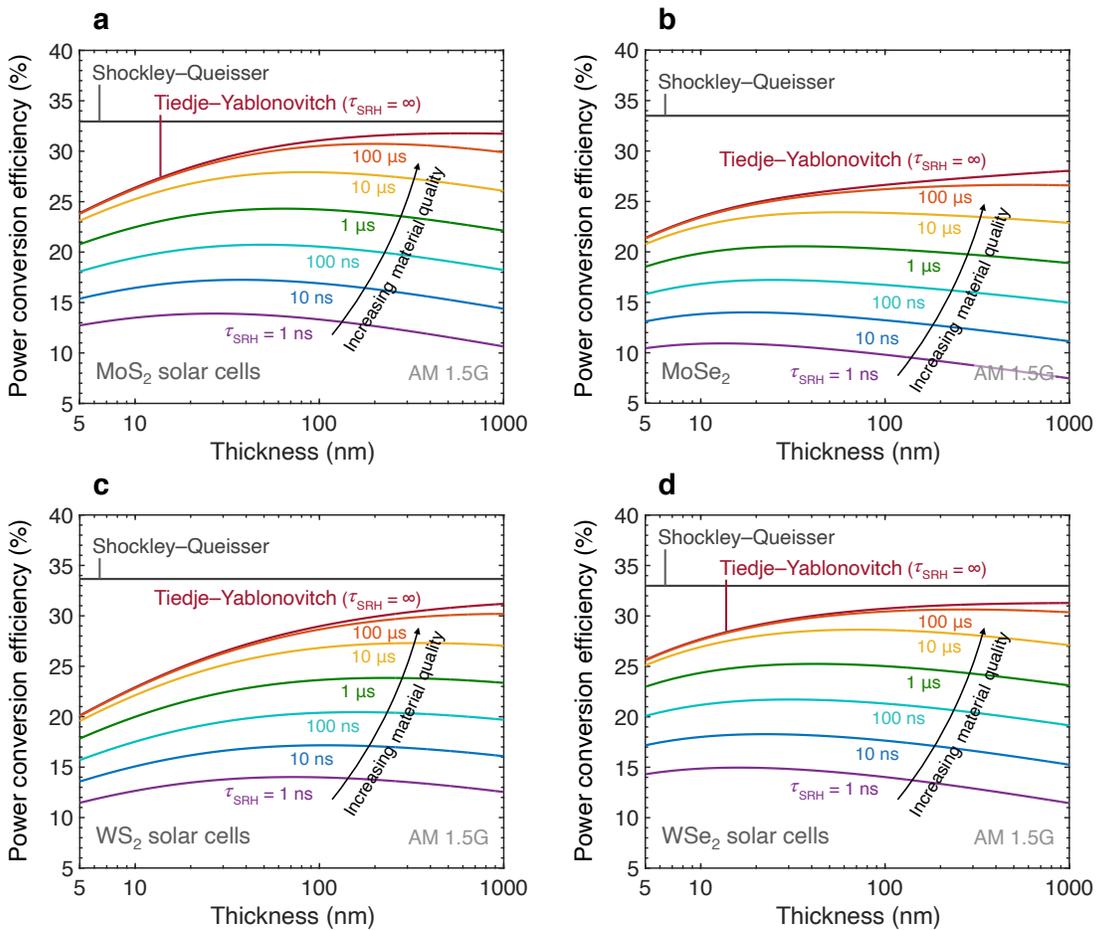

**Figure 7. Power conversion efficiency** of **a)** $MoS_2$, **b)** $MoSe_2$, **c)** $WS_2$, and **d)** $WSe_2$ solar cells as a function of TMD film thickness and material quality (SRH lifetime, $\tau_{SRH}$), at 300 K and AM 1.5 G solar illumination.

Most importantly, **Fig. 7** shows the power conversion efficiency of $MoS_2$, $MoSe_2$, $WS_2$, and $WSe_2$ solar cells as a function of TMD film thickness and material quality (i.e. SRH lifetime, $\tau_{SRH}$). The Shockley-Queisser efficiency limits are included for comparison. Given that efficiency is equal to the product of $J_{SC}$, $V_{OC}$, and fill factor, the efficiency trends observed in **Fig. 7** can be easily explained by $J_{SC}$, $V_{OC}$, and fill factor trends in **Fig. 3**, **Fig. 5,** and **Fig. 6**, respectively. As the TMD film thickness increases, $J_{SC}$ improves (**Fig. 3**), whereas both $V_{OC}$ (**Fig. 5**) and fill factor (**Fig. 6**) degrade. This competition causes the inverted U-shaped curves in **Fig. 7**, where efficiency initially increases with thickness and then decreases after a certain point. With the Tiedje-Yablonovitch model ($\tau_{SRH} \rightarrow \infty$), the maximum efficiency occurs for thicknesses over 1000 nm, therefore we only observe an increasing trend within the range of thicknesses considered here. At 100 nm absorber layer thickness, TMD solar cells achieve up to ~31% Tiedje-Yablonovitch efficiency (**Supplementary Table 1**), which is ~5% higher than Tiedje-Yablonovitch efficiency limit of silicon solar cells (29.8%) with 1000 times thicker absorber layers (100 μm). This highlights the considerable potential of TMD solar cells for ultrathin photovoltaics with high power per weight.

Going beyond the Tiedje-Yablonovitch model, we introduce non-negligible SRH recombination (i.e. reduced $\tau_{SRH}$, corresponding to reduced material quality), observing how the efficiency drops in **Fig. 7**, as a consequence of $V_{OC}$ and fill factor degradation (**Fig. 5** and **Fig. 6**). Moreover, we note that for smaller $\tau_{SRH}$ the maximum efficiency in **Fig. 7** occurs at smaller thicknesses since stronger SRH recombination leads to steeper degradation in $V_{OC}$ and fill factor with increasing film thickness. In other words, although the peak efficiency is reduced, one benefit of "more defective" TMD materials is that their efficiency is maximized in a thinner material, which could potentially have higher specific power and lower cost.

Another way to visualize the effect of material quality ($\tau_{SRH}$) on the solar cell performance is to look at current density–voltage (*J*–*V*) characteristics for a fixed thickness, for example 100 nm (**Supplementary Fig. 3**). As noted previously, within the thickness and $\tau_{SRH}$ ranges considered here, SRH recombination does not influence $J_{SC}$ due to the low carrier density at zero bias in the intrinsic or lightly-doped TMDs assumed, but it impacts both $V_{OC}$ and fill factor, therefore power conversion efficiency. We also examine the effect of Auger recombination on power conversion efficiency, varying the Auger coefficients of TMDs in the absence of SRH recombination by four orders of magnitude (**Supplementary Fig. 4**), two orders of magnitude below and above the primary Auger coefficients used in this study, which were extrapolated from Auger coefficient–band gap charts in the literature[18]. We observe that such large variation in Auger coefficients leads to a mere 1-2% change in power conversion efficiency, demonstrating the robustness of the efficiency limits modeled in this study despite the uncertainty over the exact Auger coefficient values.

The relative efficiencies of the four TMDs can be explained by their relative $J_{SC}$, $V_{OC}$, and fill factors. In the 100 ns–1 μs SRH lifetime regime, $WSe_2$ solar cells demonstrate the highest efficiency, followed by $MoS_2$, $WS_2$ and $MoSe_2$ solar cells. To date, SRH lifetimes up to 611 ns are reported in the literature for multilayer TMDs[3], corresponding to approximately 20-25% power conversion efficiency for the TMD solar cells examined here with ultrathin films of 20 to 100 nm thickness (**Fig. 7**). Such power conversion efficiency can be practically achieved by optimizing the optical and electrical design of the ultrathin TMD solar cells, yielding approximately 10 times higher specific power than existing solar cell technologies on the market[4]. As a result, such ultra-lightweight TMD solar cells could create unprecedented opportunities across various industries from aerospace to wearable electronics.

Finally, **Fig. 8** summarizes the relative magnitudes of various loss mechanisms in optimized 100 nm-thick TMD solar cells at the maximum power point (MPP), as detailed in **Supplementary Note 1**. $\tau_{SRH}$ of 100 μs is considered, where SRH recombination has comparable carrier lifetime and therefore magnitude with Auger and radiative losses (**Supplementary Table 2**). At shorter $\tau_{SRH}$, SRH recombination dominates and accounts for nearly all the recombination loss, as can be seen in **Fig. 5**, **Fig. 6** and **Fig. 7**. At a fixed $\tau_{SRH}$, the relative magnitudes of SRH recombination in various TMDs depend on their carrier densities (**Supplementary Note 1**). At their maximum power point, 100 nm-thick $MoS_2$ and $WS_2$ have the highest and lowest carrier densities (**Supplementary Table 2**), and therefore the largest and smallest current loss due to the SRH recombination.

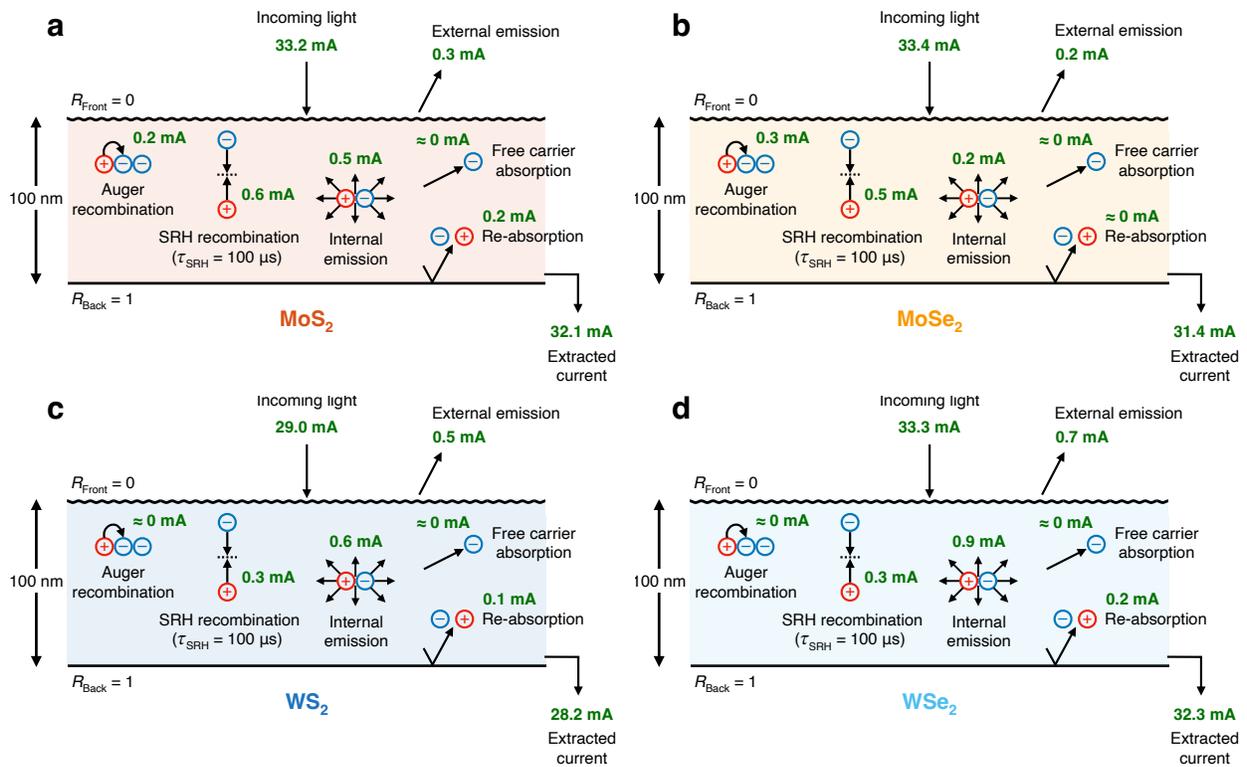

**Figure 8. Loss mechanisms at play.** Summary of relative magnitudes of various loss mechanisms in 100 nm-thick **a)** $MoS_2$, **b)** $MoSe_2$, **c)** $WS_2$, and **d)** $WSe_2$ solar cells at the maximum power point (MPP), 300 K temperature and AM 1.5 G solar illumination. SRH lifetime, $\tau_{SRH}$ = 100 μs is considered. The solar cells are assumed to have optimized electrical and optical design. Free carrier absorption is negligible given the low doping density and small thickness of the TMDs assumed here[20]. $R$, reflection.

Internal emission scales exponentially with the output voltage (**Supplementary Note 1**). As a result, WS$_2$ which has the highest band gap and therefore $V_{MPP}$ (**Supplementary Table 2**), shows the largest internal emission. For each TMD, the relative magnitudes of reabsorption and external emission are the same as in **Fig. 4**. Auger loss is proportional to Auger coefficient and the cube of carrier density at the maximum power point (**Supplementary Note 1**), therefore smallest for WS$_2$ and WSe$_2$ (see **Table I** and **Supplementary Table 2**). As discussed previously, free carrier absorption is negligible due to the low doping density and small thickness of the TMDs assumed here[20]. The relative magnitudes of SRH, Auger and radiative recombination in each TMD can be explained by their relative carrier lifetimes, given the inverse proportionality between the recombination rate and the carrier lifetime (**Supplementary Note 1**). For example, WS$_2$, with radiative and Auger lifetimes of ~50 μs and ~2 ms, respectively (**Supplementary Table 2**), exhibits ~2× larger radiative emission than SRH recombination, and negligible Auger loss.

## IV. CONCLUSIONS

We have examined the efficiency limits of multilayer TMD solar cells (MoS$_2$, MoSe$_2$, WS$_2$, and WSe$_2$) as a function of TMD film thickness and material quality, going beyond the Tiedje-Yablonovitch and Shockley-Queisser models by including experimental optical absorption spectra, as well as radiative, Auger and SRH recombination. We find that ultrathin TMD solar cells (as thin as 50 nm) can realistically achieve up to 25% power conversion efficiency even with today's material quality. This makes them an excellent choice for high-specific-power photovoltaics (i.e. with high power per weight), achieving up to 10 times higher specific power than existing technologies. Such ultralight solar cells could transform energy harvesting across various industries including autonomous drones, electric vehicles, wearable electrons, and the Internet of Things. Future efforts must be dedicated to optimizing the electronic and optical TMD solar cell designs, to unlock their potential for high power conversion efficiency and specific power at large, industrial scale.

**Data Availability:** The data that support the findings of this study are available from the corresponding author upon reasonable request.

**Authors Contributions:** K.N. and F.U.N. contributed equally. K.N. conceived the project. K.N. and F.U.N. developed the extended detailed balance model. F.U.N. implemented the model on TMDs, assisted by K.N. All authors, i.e. K.N., F.U.N., A.D., K.C.S, and E.P., contributed to the data interpretation, presentation, and writing of the manuscript. E.P. supervised the work.

**Acknowledgements:** The authors acknowledge partial support from Stanford Precourt Institute for Energy and the member companies of the SystemX Alliance at Stanford.

**Competing Interests:** The authors declare no competing interests.

# Supplementary Information

# Efficiency Limit of Transition Metal Dichalcogenide Solar Cells

Koosha Nassiri Nazif,[1†] Frederick U. Nitta,[1,2†] Alwin Daus,[1,3] Krishna C. Saraswat,[1,2] and Eric Pop[1,2*]

[1]Dept. of Electrical Engineering, Stanford University, Stanford, CA 94305, USA

[2]Dept. of Materials Science and Engineering, Stanford University, Stanford, CA 94305, USA

[3]RWTH Aachen University, Aachen, 52074, Germany

[†]These authors contributed equally.

*corresponding author email: epop@stanford.edu

**Supplementary Note 1. Extended detailed balance method considering radiative, Auger, SRH recombination, and free carrier absorption**

According to the Tiedje-Yablonovitch model[13], in the presence of radiative emission, Auger recombination and free carrier absorption, the detailed balance equation governing the current density–voltage ($J$–$V$) characteristics of an optimized solar cell having an intrinsic or lightly-doped absorber film, i.e. equal electron ($N$) and hole density ($P$) under illumination, is the following:

$$\left(\alpha_1 + \frac{1}{4n^2L}\right)\exp\left(\frac{eV}{kT}\right)\int\int a_2(E)b_n(E,T)dE\,d\Omega + CN^3 = \frac{J_{in}}{eL}(1-f) \quad (1)$$

where $\alpha_1$ is free carrier absorption coefficient, $n$ is the refractive index of the absorber film, $L$ is the thickness of the film, $e$ is the elementary charge, $V$ is the output voltage, $k$ is the Boltzmann constant, $T$ is temperature, $a_2(E)$ is absorptance (absorption probability) at photon energy $E$, $b_n(E,T)dEd\Omega$ is flux of black-body photons for a photon energy interval $dE$ and solid angle $d\Omega$ in a medium with refractive index of $n$, $C$ is Auger coefficient, $N$ is electron (and hole) density, $\frac{J_{in}}{eL}$ is the volume rate of generation of electron-hole pairs by the sun, and $f$ is fraction of the incident solar flux that is drawn off as current into the external circuit. $a_2(E)$, $b_n(E,T)$ and $J_{in}$ are defined as:

$$a_2(E) = \frac{\alpha_2(E)}{\alpha_2(E) + \alpha_1(E) + \frac{1}{4n^2L}} \quad (2)$$

$$b_n(E,T) = \frac{2}{h^3}\frac{n^2}{c^2}E^2\exp\left(\frac{1}{\frac{E}{kT}-1}\right) \quad (3)$$

$$J_{in} = \int eS(E)a_2(E)dE \quad (4)$$

where $\alpha_2(E)$ is the optical absorption coefficient at photon energy $E$, $h$ is the Planck constant, $c$ is the speed of light in vacuum, and $S(E)$ is the solar spectrum (AM 1.5 G illumination, one-sun intensity). The left-hand side of **Equation (1)**, from left to right, refers to the rate of free carrier absorption, radiative emission, and Auger recombination, while the right-hand side, from left to right, refers to photogenerated electron-hole pairs and the output current of the solar cell. To include SRH recombination, we add SRH recombination rate $U_{SRH}$ to the left-hand side of **Equation (1)**:

$$\left(\alpha_1 + \frac{1}{4n^2L}\right)\exp\left(\frac{eV}{kT}\right)\int\int a_2(E)b_n(E,T)dE\,d\Omega + CN^3 + U_{SRH} = \frac{J_{in}}{eL}(1-f) \quad (5)$$

For any recombination mechanism, associated carrier lifetimes, $\tau_e$ and $\tau_p$, for electrons and holes, can be defined as:

$$\tau_e = \frac{\Delta N}{U} \tag{6}$$

$$\tau_p = \frac{\Delta P}{U} \tag{7}$$

where $\Delta N$ and $\Delta P$ are the disturbances of the electrons and holes, respectively, from their equilibrium values $N_0$ and $P_0$. $U$ is the recombination rate. For an intrinsic or lightly-doped absorber film under illumination:

$$N = P \gg N_0, P_0 \tag{8}$$

$$\Delta N = \Delta P \approx N \tag{9}$$

Therefore the SRH recombination rate can be written as:

$$U_{SRH} = \frac{N}{\tau_{SRH}} \tag{10}$$

Combining **Equations (5)** and **(10)** leads to the following:

$$\left(\alpha_1 + \frac{1}{4n^2 L}\right) \exp\left(\frac{eV}{kT}\right) \int \int a_2(E) b_n(E,T) dE\, d\Omega + CN^3 + \frac{N}{\tau_{SRH}} = \frac{J_{in}}{eL}(1-f) \tag{11}$$

**Equation (11)** is the detailed balance equation governing the current density–voltage characteristics of an optimized solar cell having an intrinsic or lightly-doped absorber film (i.e., N = P under illumination) in the presence radiative emission, Auger recombination, free carrier absorption, and SRH recombination with the characteristic carrier lifetime of $\tau_{SRH}$. In the absence of free carrier absorption, **Equation (11)** simplifies to the following:

$$J_0 \exp\left(\frac{eV}{kT}\right) + eLCN_i^3 \exp\left(\frac{3eV}{2kT}\right) + \frac{eL}{\tau_{SRH}} N_i \exp\left(\frac{eV}{2kT}\right) = J_{in}(1-f) \tag{12}$$

where $N_i$ is the intrinsic carrier density and $J_0$ is defined as:

$$J_0 = e\pi \int b_1(E) a_2(E) dE \tag{13}$$

Note that all terms in **Equation (12)**, i.e., external emission, Auger recombination, SRH recombination, incoming sunlight and extracted electricity (going from left to right), are all expressed in the form of current density, and are plotted in **Fig. 8** for comparison.

To obtain the current density–voltage characteristics of the solar cell, $f$ is varied from zero to one – corresponding to output current density ($J$) of zero to $J_{in}$ – and the output voltage ($V$) is subsequently determined by solving **Equation (12).** Performance metrics are extracted from the resulting $J$–$V$ characteristics, as follows:

$$V_{OC} = V(J = 0) \tag{14}$$

$$J_{SC} = J(V = 0) \tag{15}$$

$$P_{MPP} = \max(I \cdot V) = I \cdot V(\frac{d(I \cdot V)}{dV} = 0) \tag{16}$$

$$FF = \frac{P_{MPP}}{V_{OC} \cdot J_{SC}} \tag{17}$$

$$PCE = \frac{P_{MPP}}{P_{in}} = \frac{P_{MPP}}{100 \; W \; cm^{-2}} \tag{18}$$

where $V_{OC}$ is the open-circuit voltage, $J_{SC}$ is the short-circuit current density, $P_{MPP}$ is power density at maximum power point (MPP), $FF$ is the fill factor, $P_{in}$ is the input solar power density, and $PCE$ is power conversion efficiency of the solar cell.

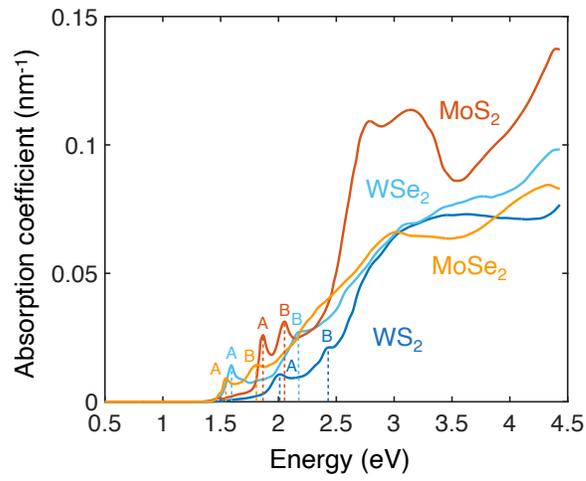

**Supplementary Figure 1 | Spectral absorption coefficient**[16] of $MoS_2$, $MoSe_2$, $WS_2$, and $WSe_2$. Excitonic peaks corresponding to A and B excitons[25] are highlighted with letters A and B.

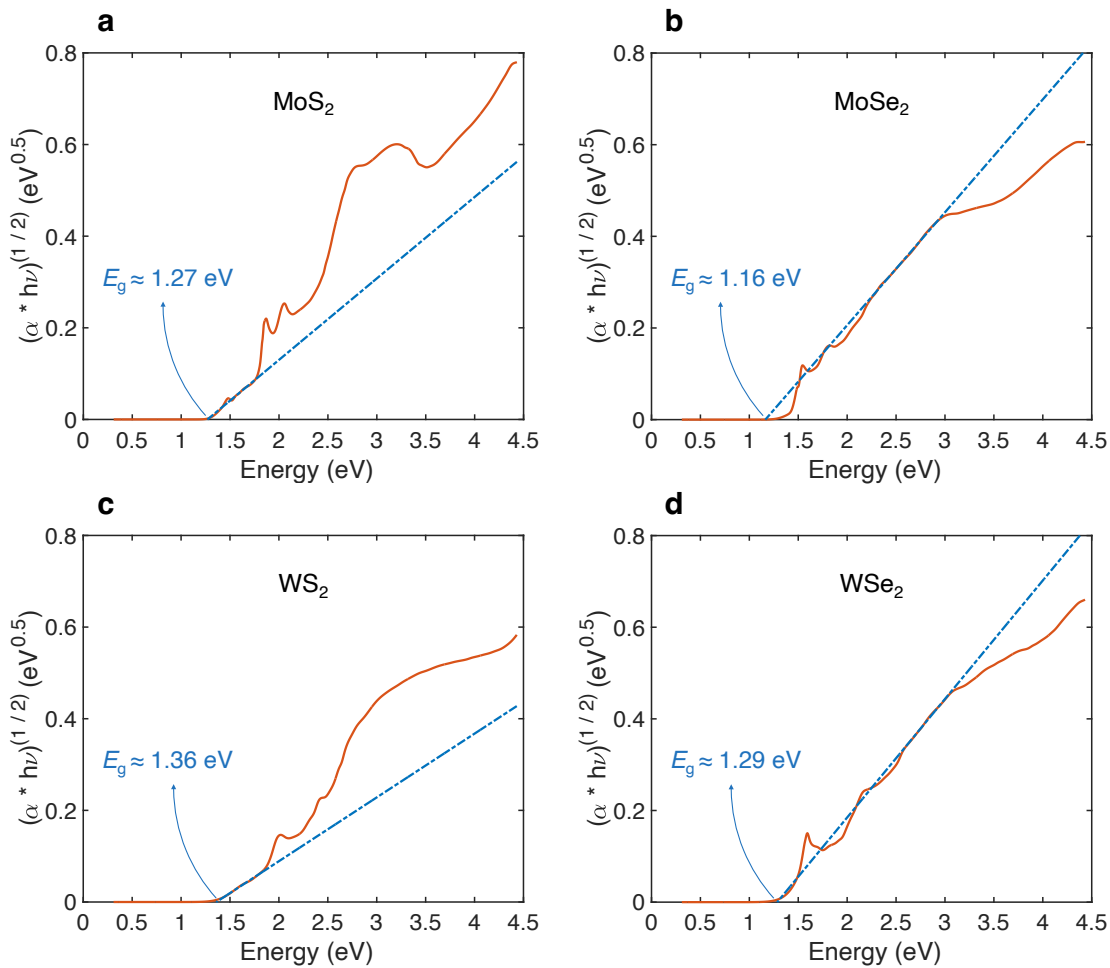

**Supplementary Figure 2 | Tauc plots** of **a)** $MoS_2$, **b)** $MoSe_2$, **c)** $WS_2$, and **d)** $WSe_2$, to determine their optical band gaps from the corresponding absorption spectra[19]. $\alpha$, absorption coefficient. $h\nu$, photon energy

**Supplementary Note 2. Luminescent emission rates**

It can be shown through detailed balance arguments[13] that the volume rate of external emission, i.e., radiation from the electron-hole pairs within a semiconductor at temperature $T$ in equilibrium with an external thermal bath is:

$$R_{external} = \frac{1}{4L} a_2(E) b_1(E,T) \tag{19}$$

Where $L$ is the thickness of the film, $a_2(E)$ is absorptance of the semiconductor, and $b_1(E,T)$ is the blackbody spectral radiance in a medium with refractive index of one ($n = 1$, air), both defined in **Supplementary Note 1.** Similarly, the rate of internal emission, i.e., radiation from the semiconductor to an internal black body (itself) with a refractive index of $n$, at equilibrium, is the following:

$$R_{internal} = a_2(E) b_n(E,T) = n^2 a_2(E) b_1(E,T) \tag{20}$$

And finally, the rate of reabsorption of internal emission, at equilibrium, is given by:

$$R_{reabsorption} = \alpha_2 a_2(E) b_n(E,T) = n^2 \alpha_2 a_2(E) b_1(E,T) \tag{21}$$

where $\alpha_2$ is the optical absorption coefficient of the semiconductor. In the absence of free carrier absorption, the following holds:

$$R_{internal} = R_{external} + R_{reabsorption} \tag{22}$$

**Supplementary Table 1 | Performance limits of 100 nm-thick MoS$_2$, MoSe$_2$, WS$_2$ and WSe$_2$ solar cells, calculated using the Tiedje-Yablonovitch model.** These ultrathin TMD solar cells exhibit Tiedje-Yablonovitch efficiencies as high as 31%. $J_{SC}$, short-circuit current density; $V_{OC}$, open-circuit voltage; FF, fill factor; PCE, power conversion efficiency, MPP, maximum power point.

|  | MoS$_2$ | MoSe$_2$ | WS$_2$ | WSe$_2$ |
|---|---|---|---|---|
| Thickness (nm) | 100 | 100 | 100 | 100 |
| $J_{SC}$ (mA cm$^{-2}$) | 33.2 | 32.5 | 29.1 | 33.3 |
| $V_{OC}$ (V) | 1.04 | 0.91 | 1.12 | 1.04 |
| Fill factor | 0.91 | 0.90 | 0.90 | 0.89 |
| $V_{MPP}$ (mV) | 960 | 842 | 1022 | 947 |
| $J_{MPP}$ (mA cm$^{-2}$) | 32.4 | 31.7 | 28.4 | 32.5 |
| Power conversion efficiency (%) | 31.1 | 26.7 | 29.0 | 30.8 |

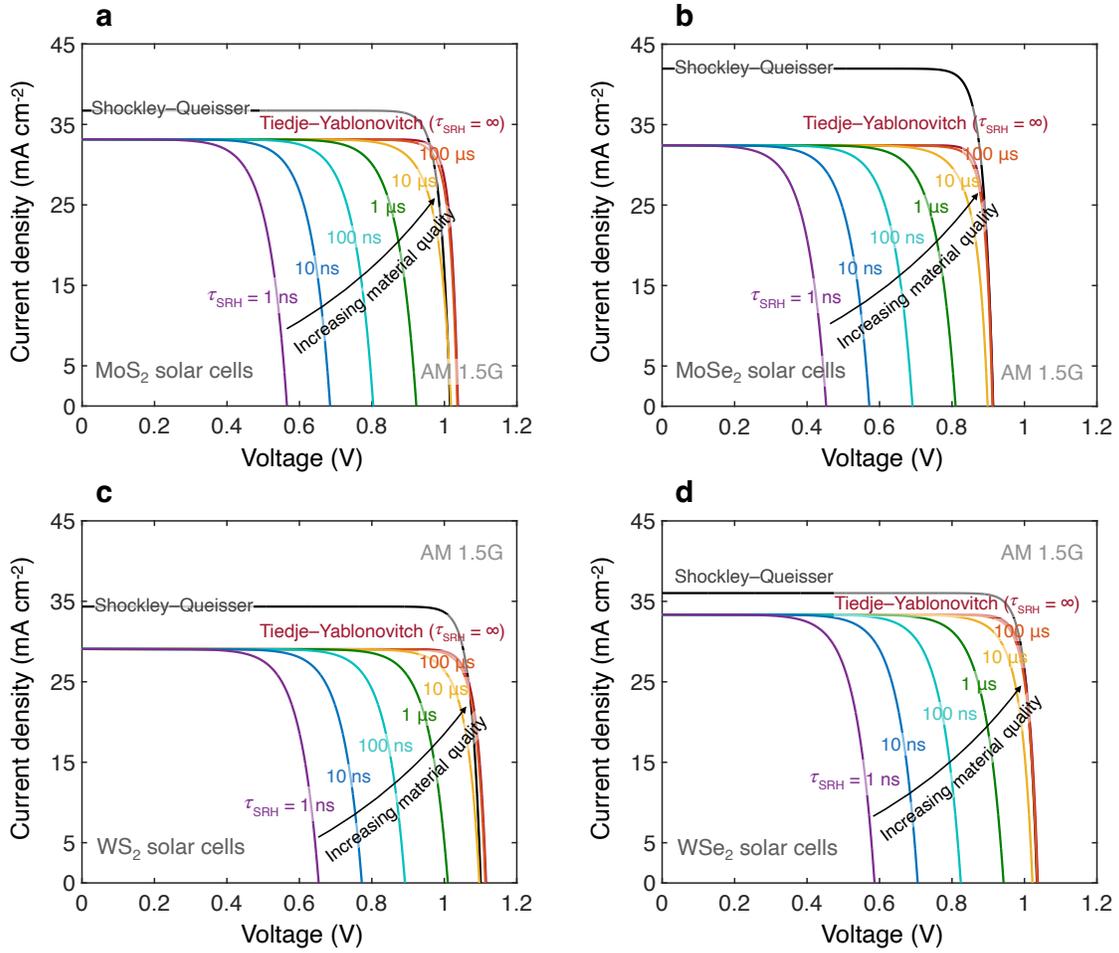

**Supplementary Figure 3 | *J–V* characteristics of 100 nm-thick a)** $MoS_2$, **b)** $MoSe_2$, **c)** $WS_2$, and **d)** $WSe_2$ solar cells as a function of material quality (SRH lifetime, $\tau_{SRH}$), at 300 K and AM 1.5 G solar illumination. Shockley-Queisser *J–V* characteristics are included for reference.

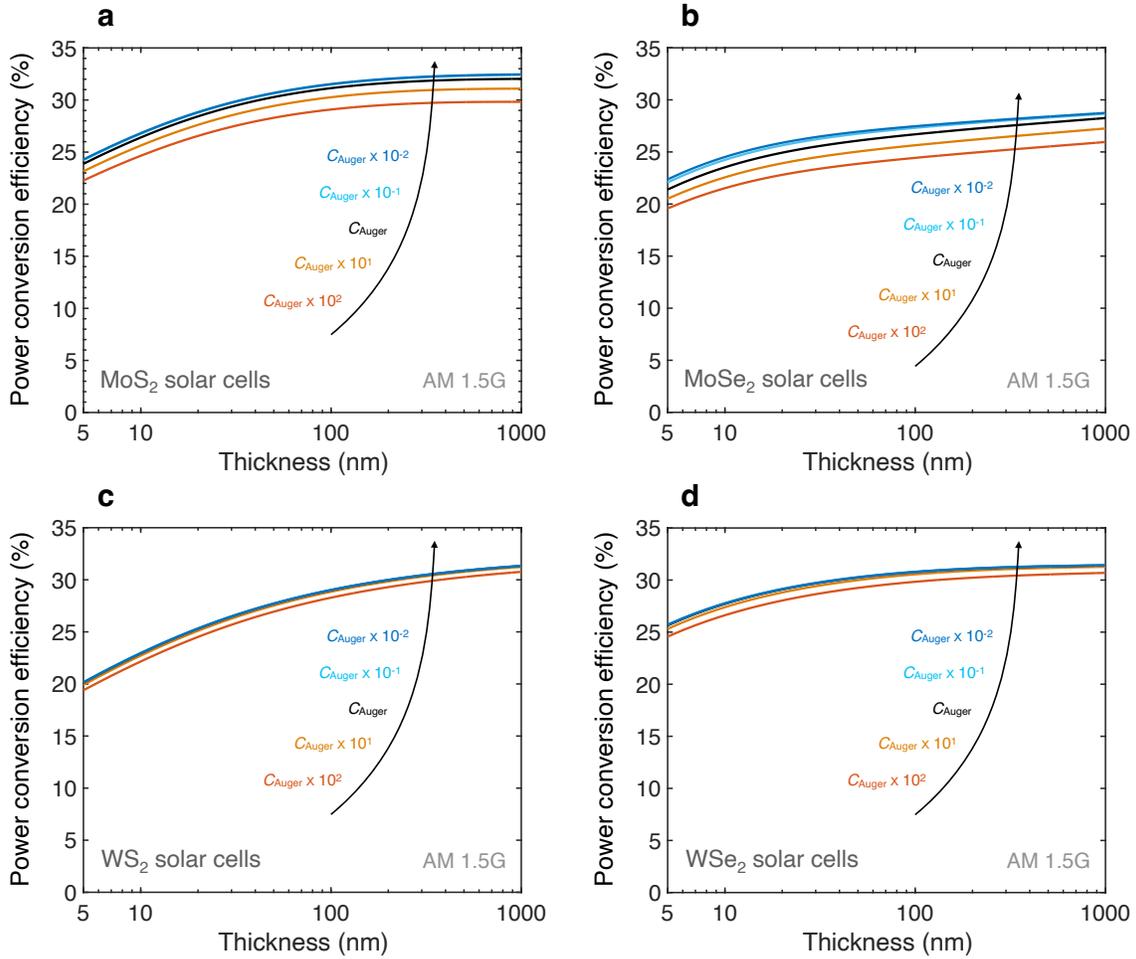

**Supplementary Figure 4 | Effect of Auger coefficient on power conversion efficiency.** Power conversion efficiency of **a)** $MoS_2$, **b)** $MoSe_2$, **c)** $WS_2$, and **d)** $WSe_2$ solar cells as a function of thickness and Auger coefficient, in the absence of SRH recombination. $C_{Auger}$ is the Auger coefficient used in this study. The figure shows that two orders of magnitude higher or smaller $C_{Auger}$ lead to at most 1-2% decrease or increase in the power conversion efficiency limit.